# CSHURI – Modified HURI algorithm for Customer Segmentation and Transaction Profitability

Jyothi Pillai [1] and O.P.Vyas [2]

[1] Associate Professor, Bhilai Institute Of Technology, Durg, Chhattisgarh, India
jyothi_rpillai@rediffmail.com
[2] Professor, Indian Institute of Information Technology Allahabad, U.P., India
dropvyas@gmail.com

## ABSTRACT

*Association rule mining (ARM) is the process of generating rules based on the correlation between the set of items that the customers purchase.Of late, data mining researchers have improved upon the quality of association rule mining for business development by incorporating factors like value (utility), quantity of items sold (weight) and profit. The rules mined without considering utility values (profit margin) will lead to a probable loss of profitable rules.*

*The advantage of wealth of the customers' needs information and rules aids the retailer in designing his store layout[9]. An algorithm CSHURI, Customer Segmentation using HURI, is proposed, a modified version of HURI [6], finds customers who purchase high profitable rare items and accordingly classify the customers based on some criteria; for example, a retail business may need to identify valuable customers who are major contributors to a company's overall profit. For a potential customer arriving in the store, which customer group one should belong to according to customer needs, what are the preferred functional features or products that the customer focuses on and what kind of offers will satisfy the customer, etc., finds the key in targeting customers to improve sales [9], which forms the base for customer utility mining.*

## 1. INTRODUCTION

Many business enterprises accumulate large quantities of data from day-to-day operations. For example, huge amount of customer purchase data are collected daily at the checkout counters or grocery stores. Retailers are interested in analyzing the data to learn about the purchasing behaviour of the customers. Such valuable information can be used to support a variety of business related applications such as marketing promotions, inventory management and customer relation management [10].

Data mining techniques can be used to support a wide range of business intelligence applications such as customer profiling, targeted marketing, work flow management, store layout and fraud detection. It can also help retailers answer important business questions such as the most important customers or most profitable transactions, etc. [10]

Over the years Data Mining is used to understand the consumer buying behaviour using various techniques. Researcher selects Market Basket Analysis for his data analysis because Market Basket analysis is a tool of knowledge discovery about co-occurrence of nominal or categorical items. Market Basket Transaction or market Basket Analysis is a data mining technique to derive association between data sets. The discovery of interesting association relationship among huge amounts of customer transaction records can help in many business decision making processes such as catalog design, cross marketing and loss-leader analysis[2].

The association analysis described so far is based on the pretext that the presence of an item in a transaction is more important than its absence. As a consequence patterns that are rarely found in a database are often considered to be uninteresting and are eliminated using the support measure. Such patterns are known as infrequent patterns. An infrequent pattern is an itemset or a rule whose support is less than the minimum support threshold [10]. In most business applications, frequent itemsets may not generate much profit. Rare itemsets are very important and can be further promoted together because they possess high associations and can bring some acceptable profits. For major expansion drive, the stores can shortlist high profitable items and can increase its quantity and thereby can earn profit.

Point-of-sale data collection (bar code scanners, radio-frequency identification and smartcard technology) have allowed retailers to collect up-to-the minute data about customer purchases at the checkout counters of their stores. Retailers can utilize this information, along with other business critical data such as Weblogs from e-commerce web sites and customer service records from call centres, to help them better understand the needs of their customers and make more informed business decisions [10].

It is proposed that Customer Utility Mining, using the CSHURI algorithm (**C**ustomer **S**egmentation using **HURI**), finds customers who purchase high profitable rare items and accordingly classify customers for providing good customers' service. After finding such customers, they can be categorized and accordingly gold or silver cards can be issued for these types of customers or a new scheme can be introduced for them. These customers provide maximum profit to the overall transaction scenario. Stores can use its customer base as a bargaining power to strike discount deals or can gain wholesale trust of consumers in a very short span. The outcome of CSHURI would enable the top management or business analyst in crucial decision-making such as catalog design, providing credit facility, cross marketing, finalizing discount policy, analyzing consumers' buying behaviour, organizing shelf space, loss-leader analysis and quality improvement in supermarket.

In this paper, it is proposed that CSHURI can be used both for Customer Utility Mining and Transaction Utility Mining for classifying customers and for finding profitable transactions containing rare itemsets. The rest of paper is organized as follows. Section 2 presents some related works. Section 3 and section 4 discuss theoretical definitions and proposed CSHURI algorithm; the modified version of HURI algorithm Section 5 presents conclusion and future work.

## 2. LITERATURE SURVEY

The problem of utility-based itemset mining is to discover the itemsets that are significant according to their utility values. In [13], Yao et al propose a utility-based itemset mining approach which permits users to quantify their preferences concerning the usefulness of itemsets using utility values. Two algorithms UMining and UMining_H were proposed, for utility-based itemset mining by incorporating pruning strategies. UMining guarantees that all high utility itemsets are found, while the heuristic share based methods, UMining_H, may miss many relevant itemsets.

David et al presented a new algorithm, MINIT ( MINimal Infrequent iTemsets), for finding minimal infrequent or minimal occurrent itemsets. Initially, a ranking of items is prepared by computing the support of each of the items and then creating a list of items in ascending order of support. Minimal infrequent itemsets are discovered by considering each item $i_j$ in rank order, recursively calling MINIT on the support set of the dataset with respect to $i_j$ considering only those items with higher rank than $i_j$, and then checking each candidate MII against the original dataset[11].

In the Frequent Itemset Mining problem, the occurrence of each item in a transaction is represented by a binary value without considering its quantity or an associated weight such as price or profit. However, quantity and weight are significant for addressing real world decision problems that require maximizing the utility in an organization. For example, selling a laser printer may occur less frequently than sale of printer ink in an electronic superstore, but the former gives a much higher profit per unit sold. The high utility itemset mining problem is to find all itemsets that have utility larger than a user specified value of minimum utility. In A Bottom-Up Projection Based Algorithm for Mining High Utility Itemsets [12], Alva et al proposed the *CTU-PRO* algorithm to mine the complete set of high utility itemsets from both sparse and relatively dense datasets with short or long high utility patterns.

In the article **Data Mining: A Tool for the Enhancement of Banking Sector** [1], Shipra et al described that data mining can be a very powerful and helpful tool to extract important and useful information for banking sector from the historical as well as from the current data. Data mining can be used in various fields of banking like Market segmentation by which banks can segment their customers into different groups, direct mail marketing can help the banks to improve their marketing strategy and to increase their business, customer churn to increase the rate of retention of the customers, risk management to reduce the various risks like creditworthiness and fraud detection to reduce the number of fraudulent.

In [7], the authors have presented a novel utility FP-tree by utilizing a tree structure for storing essential information about frequent patterns for mining high utility itemsets. Higher efficiency in mining high utility patterns is realized by implementing two important concepts. One is the construction of the utility FP-tree and the other one is the mining of utility itemsets from the utility FP-tree. The utility FP-tree-based pattern mining utilizes the pattern growth method to avoid the costly generation of a large number of candidate sets and reduces the search space dramatically. The experimentation was carried out using real life datasets and the results show that the proposed approach effectively mines utility itemsets from large databases.

Jyothi et al proposed a High Utility Rare Itemset Mining [HURI] algorithm [6], to find those rare itemsets, which are of high utility according to users' preferences. Using HURI algorithm, high utility rare itemsets are generated in two phases:-
(i)In first phase, rare itemsets are generated by considering those itemsets which have support value less than the maximum support threshold.
(ii)In second phase, by inputting the utility threshold value according to users' interest, rare itemsets having utility value greater than the minimum utility threshold are generated.
HURI can produce high utility rare itemsets based on support threshold, utility threshold and users' interest.

# 3. PROBLEM DEFINITION

First theoretical related concepts of the proposed algorithm, CSHURI, are described.

**DEFINITION 3.1 (Utility Mining)** Utility Mining finds all itemsets in transaction database with utility values higher than the user defined minimum utility threshold.

Let I be a set of quantities of items I={$i_1$, $i_2$, $i_3$,… , $i_m$}and  D be a set of transactions {$T_1$,$T_2$,…,$T_n$} with items, where each item i ε I(table 1). Each transaction in D is assigned a transaction identifier (T_ID). The set of utilities is defined as U={$u_1$, $u_2$, $u_3$,… , $u_k$} (table 2).  For e.g. in transaction $T_{19}$, the quantities of items A, B, C, D, E… are 0,0,0,1,2,… respectively.

The utility of an itemset *X*, i.e., *u(X)*, is the sum of the utilities of itemset X in all the transactions containing *X*. An itemset *X* is called a *high utility itemset* if and only if *u(X)* >= *min_utility*, where *min_utility* is a user-defined minimum utility threshold [14]. Identification of the itemsets with high utilities is called as Utility Mining [15].

**DEFINITION 3.2 (Utility Table)** A utility table UT (table 2) is a table containing items and their corresponding utility values where each item i has some utility value $u_j$ in U={$u_1$, $u_2$, $u_3$,… , $u_k$ } for some k > 0.

For example utility of item E is u(E) = 7 in (table 2).

**DEFINITION 3.3 (Internal Utility)** The internal utility value of item $i_p$ in a transaction $T_q$, denoted o($i_p$, $T_q$) is the value of an item $i_p$ in a transaction $T_q$ (Table 2). The internal utility reflects the occurrence of the item in a transaction database.

 In table 1, internal utility of item A in transaction T1 is o(A, T1) = 1, while internal utility of item A  in Transaction dataset D is o(A, D) = 21.

**DEFINITION 3.4 (External Utility)** The external utility value of an item is a numerical value s($i_p$) associated with an item $i_p$ such that s($i_p$ )=u($i_p$), where u is a utility function, a function relating specific values in a domain according to user preferences (table 2).

From table 3, external utility of item A is s(A) = u(A) = 4.

**DEFINITION 3.5 (Item Utility)** The utility of an item $i_p$ in a transaction $T_q$, denoted U($i_p$, $T_q$) is product of o($i_p$, $T_q$) and s($i_p$), where o($i_p$, $T_q$) is the internal utility value of $i_p$, s($i_p$) is the external utility value of $i_p$(table 3) .

For eg., total utility of  item A  is U(A) = s(A)  * o(A) = 4 * 21 = 84 (table 2).

**DEFINITION 3.6 (Transaction Utility)** The transaction utility value of a transaction, denoted as U($T_q$) is the sum of utility values of all items in a transaction $T_q$  (table 1, table 2). The transaction utility reflects the utility in a transaction database.

From Table 1 and Table 3, the transaction utility of the transaction T1, U(T1) = U(A)+U(B)+U(C)+U(D)+ … + U(T)=39.

**DEFINITION 3.7 (Rare Itemset Mining)** Rare itemsets are the itemsets that occur infrequently in the transaction data set. In many practical situations, the rare combinations of items in the itemset with high utilities provide very useful insights to the user. Some infrequent patterns may also suggest the occurrence of interesting rare events or exceptional situations in the data. For e.g. If {Fire=Yes} is frequent but {Fire=Yes, Alarm=ON} is infrequent, then latter is an interesting infrequent pattern because it may indicate faulty alarm system [10].

Rare itemset mining is a challenging task. The key issues in mining rare itemsets are: -
  (i)   How to identify interesting rare patterns and
  (ii)  How to efficiently discover them in large datasets.

# 4. Proposed Algorithm

In [6], authors propose a High Utility Rare Itemset Mining [HURI] algorithm for generating high utility rare itemsets of users' interest. One more very interesting and innovative idea is to use HURI as a base for customer utility mining. The proposed CSHURI algorithm, **C**ustomer **S**egmentation using **HURI,** finds customers who purchase high profitable rare items and accordingly classify the customers for providing good customers' service.

**Algorithm CSHURI**

**Description: Finding High Utility Rare Itemsets of users' interest and classifying customers**
$C_k$: **Candidate itemset of size k**
$L_k$: **Rare itemset of size k**

For each transaction $t$ in database
**begin**
   increment support for each item $i$ present in $t$
**End**

$L_1$= {Rare 1-itemset with support less than user provided max_sup}
**for($k$= 1; $L_k$!=Ø; $k$++)**
**begin**
   $C_{k+1}$= candidates generated from $L_k$;
   **//loop to calculate total utility of each item**
   For each transaction $t$ in database
   **begin**
    Calculate total quantity of each item $i$ in $t$
    Find total utility for item $i$ using following formula:-
        **u(i,t) = quantity[i] * user_provided_utility for i**
   **End**

   **//loop to find rare itemsets and their utility**
   For each transaction $t$ in database
   **begin**
     increment the count of all candidates in $C_{k+1}$ that are contained in $t$
     $L_{k+1}$ = candidates in $C_{k+1}$ *less than* min_support
        Add $L_{k+1}$ to the Itemset_Utility table in database by calculating rare itemset utility using formula:
            **Utility(R,t) = $\Sigma_{\text{for each individual item i in R}}$ (u(i,t));**
   **End**

   **//loop to find high utility rare itemset**
   For each itemset $iset$ in rare itemset table $R$
   **begin**
     If (**Utility(iset)** > user_provided_threshold_for_high_utility_rare_itemset)
     then **iset** is a rare_itemset that is of user interest i.e.high_utility_rare_itemset
     else **iset** is a rare itemset but is not of user interest
   **End**

**//loop to calculate profit of each transaction and use this profit as base for customer_utility_mining**
  For each transaction *t* in database
  **begin**
  Set profit of each transaction *t* in customer utility table as
  **Profit_transaction_t = (utility of each item i) \* (quantity of item i in t)**
    If (Profit_transaction_t > user_provided_cust_utility)
        Customer is a **premium** customer
    Else
        Customer is a **general** customer
  **End**
  **Return high_utility**_rare_itemsets, premium_customers
**END**

**Figure 1 : Pseudo Code for CSHURI**

**Table 1: Transaction  Dataset D**

| TID | A | B | C | D | E | F | G | H | I | J | K | L | M | N | O | P | Q | R | S | T |
|-----|---|---|---|---|---|---|---|---|---|---|---|---|---|---|---|---|---|---|---|---|
| T1  | 1 | 2 | 2 | 0 | 0 | 1 | 1 | 0 | 2 | 0 | 1 | 0 | 5 | 0 | 0 | 1 | 4 | 0 | 1 | 0 |
| T2  | 1 | 0 | 1 | 1 | 1 | 0 | 0 | 0 | 0 | 3 | 1 | 1 | 0 | 4 | 0 | 1 | 0 | 3 | 0 | 1 |
| T3  | 1 | 3 | 2 | 0 | 0 | 0 | 0 | 0 | 2 | 0 | 1 | 0 | 3 | 2 | 1 | 0 | 0 | 0 | 0 | 1 |
| T4  | 0 | 0 | 0 | 0 | 1 | 3 | 0 | 0 | 0 | 4 | 1 | 1 | 0 | 1 | 0 | 1 | 0 | 1 | 1 | 0 |
| T5  | 0 | 1 | 0 | 0 | 1 | 0 | 1 | 0 | 1 | 0 | 0 | 0 | 1 | 0 | 1 | 0 | 1 | 1 | 0 | 0 |
| T6  | 0 | 2 | 0 | 0 | 0 | 0 | 0 | 0 | 1 | 0 | 0 | 0 | 0 | 0 | 1 | 0 | 1 | 3 | 0 | 1 |
| T7  | 0 | 0 | 0 | 0 | 0 | 0 | 0 | 0 | 1 | 0 | 0 | 1 | 0 | 0 | 1 | 1 | 1 | 5 | 1 | 1 |
| T8  | 1 | 0 | 1 | 1 | 1 | 0 | 0 | 0 | 3 | 0 | 1 | 0 | 4 | 4 | 0 | 1 | 0 | 0 | 0 | 0 |
| T9  | 0 | 0 | 1 | 0 | 2 | 4 | 0 | 2 | 0 | 0 | 0 | 1 | 0 | 0 | 1 | 1 | 0 | 4 | 1 | 1 |
| T10 | 2 | 3 | 1 | 1 | 1 | 0 | 0 | 0 | 5 | 0 | 0 | 1 | 6 | 2 | 1 | 1 | 6 | 0 | 0 | 0 |
| T11 | 1 | 1 | 0 | 0 | 0 | 1 | 0 | 0 | 0 | 0 | 3 | 1 | 0 | 0 | 0 | 0 | 0 | 3 | 0 | 1 |
| T12 | 1 | 0 | 1 | 0 | 1 | 0 | 1 | 1 | 1 | 0 | 0 | 1 | 5 | 1 | 0 | 0 | 0 | 0 | 1 | 1 |
| T13 | 0 | 2 | 0 | 1 | 0 | 0 | 0 | 1 | 3 | 1 | 1 | 0 | 0 | 0 | 1 | 1 | 1 | 1 | 1 | 0 |
| T14 | 0 | 0 | 1 | 0 | 2 | 3 | 1 | 0 | 1 | 5 | 0 | 0 | 3 | 2 | 0 | 0 | 5 | 0 | 1 | 1 |
| T15 | 1 | 1 | 1 | 0 | 1 | 0 | 0 | 0 | 1 | 1 | 1 | 1 | 0 | 0 | 1 | 1 | 1 | 2 | 0 | 1 |
| T16 | 0 | 0 | 0 | 0 | 0 | 0 | 0 | 0 | 0 | 0 | 0 | 2 | 0 | 0 | 0 | 0 | 0 | 0 | 0 | 0 |
| T17 | 0 | 0 | 0 | 0 | 0 | 0 | 0 | 2 | 1 | 1 | 1 | 0 | 0 | 4 | 0 | 1 | 0 | 2 | 0 | 1 |
| T18 | 1 | 3 | 0 | 0 | 1 | 4 | 0 | 0 | 0 | 0 | 0 | 0 | 5 | 0 | 0 | 1 | 0 | 0 | 1 | 0 |
| T19 | 0 | 0 | 0 | 1 | 2 | 0 | 0 | 1 | 0 | 0 | 0 | 1 | 0 | 2 | 0 | 1 | 1 | 1 | 0 | 1 |
| T20 | 0 | 0 | 2 | 0 | 0 | 0 | 1 | 2 | 0 | 0 | 0 | 0 | 1 | 0 | 0 | 1 | 5 | 0 | 1 | 0 |
| T21 | 2 | 0 | 1 | 0 | 0 | 3 | 0 | 1 | 0 | 2 | 0 | 1 | 1 | 1 | 0 | 0 | 0 | 3 | 0 | 1 |
| T22 | 0 | 0 | 2 | 0 | 0 | 0 | 0 | 0 | 0 | 0 | 0 | 0 | 1 | 1 | 0 | 0 | 1 | 0 | 1 | 0 |
| T23 | 0 | 0 | 0 | 0 | 2 | 1 | 1 | 0 | 1 | 0 | 0 | 0 | 1 | 0 | 0 | 0 | 1 | 2 | 1 | 1 |
| T24 | 2 | 0 | 1 | 0 | 0 | 0 | 0 | 0 | 0 | 0 | 0 | 0 | 0 | 0 | 0 | 1 | 4 | 1 | 0 | 0 |
| T25 | 2 | 2 | 1 | 1 | 1 | 0 | 1 | 0 | 1 | 1 | 2 | 1 | 4 | 1 | 0 | 0 | 1 | 1 | 0 | 0 |
| T26 | 0 | 0 | 0 | 2 | 1 | 0 | 0 | 0 | 0 | 0 | 0 | 1 | 1 | 1 | 0 | 1 | 0 | 1 | 0 | 1 |
| T27 | 1 | 0 | 0 | 0 | 0 | 0 | 1 | 0 | 1 | 0 | 0 | 1 | 5 | 0 | 0 | 2 | 5 | 0 | 1 | 1 |
| T28 | 0 | 0 | 0 | 0 | 0 | 4 | 0 | 1 | 2 | 0 | 0 | 0 | 2 | 0 | 0 | 1 | 0 | 0 | 1 | 1 |
| T29 | 1 | 3 | 0 | 1 | 1 | 2 | 1 | 0 | 1 | 2 | 1 | 0 | 0 | 1 | 0 | 2 | 2 | 0 | 0 | 0 |
| T30 | 0 | 0 | 2 | 0 | 0 | 0 | 0 | 1 | 0 | 0 | 0 | 0 | 2 | 1 | 0 | 1 | 0 | 2 | 0 | 0 |
| T31 | 2 | 1 | 0 | 1 | 1 | 0 | 1 | 0 | 1 | 0 | 0 | 0 | 2 | 1 | 0 | 1 | 0 | 2 | 0 | 1 |
| T32 | 0 | 1 | 0 | 1 | 1 | 0 | 1 | 0 | 1 | 0 | 0 | 0 | 2 | 1 | 0 | 1 | 0 | 2 | 0 | 1 |
| T33 | 1 | 1 | 0 | 1 | 0 | 1 | 1 | 0 | 1 | 1 | 1 | 1 | 5 | 2 | 0 | 0 | 1 | 1 | 0 | 0 |
| T34 | 0 | 1 | 0 | 0 | 0 | 0 | 0 | 1 | 1 | 1 | 0 | 0 | 1 | 0 | 0 | 1 | 0 | 0 | 0 | 0 |
| T35 | 0 | 1 | 1 | 1 | 1 | 2 | 0 | 0 | 0 | 0 | 1 | 1 | 1 | 1 | 0 | 0 | 1 | 1 | 1 | 0 |

**Table 2: Item Utility Table**

| Items | External Utility | Internal Utility | Total Utility |
|-------|------------------|------------------|---------------|
| A | 4 | 21 | 84 |
| B | 1 | 28 | 28 |
| C | 3 | 21 | 63 |
| D | 2 | 12 | 24 |
| E | 7 | 23 | 161 |
| F | 5 | 27 | 135 |
| G | 6 | 10 | 60 |
| H | 1 | 13 | 13 |
| I | 1 | 34 | 34 |
| J | 4 | 27 | 108 |
| K | 3 | 15 | 45 |
| L | 1 | 14 | 14 |
| M | 1 | 50 | 50 |
| N | 2 | 40 | 80 |
| O | 3 | 14 | 42 |
| P | 1 | 18 | 18 |
| Q | 1 | 42 | 42 |
| R | 1 | 44 | 44 |
| S | 1 | 11 | 11 |
| T | 0 | 17 | 0 |

**Table 3: Rare Itemset Table**

| Rare itemsets | List of rare itemsets | Itemset Utility |
|---|---|---|
| 1-itemset | {D} | 24 |
| | {G} | 60 |
| | {H} | 13 |
| | {S} | 11 |
| 2-itemset | {D,G} | 84 |
| | {D, H} | 37 |
| | {D,S} | 35 |
| | {G,H} | 73 |
| | {G,S} | 71 |
| | {H,S} | 24 |
| 3-itemset | {D,G,H} | 97 |
| | {D,G,S} | 95 |
| | {G,H,S} | 84 |
| | {D,H,S} | 48 |
| 4-itemset | {D,G, H,S} | 108 |

**Table 4: High Utility Rare Itemset Table**

| Rare itemsets | List of high utility rare itemsets | Utility |
|---|---|---|
| 1-itemset | {G} | 60 |
| 2-itemset | {D,G} | 84 |
| | {G,H} | 73 |
| | {G,S} | 71 |
| 3-itemset | {D,G,H} | 97 |
| | {D,G,S} | 95 |
| | {G,H,S} | 84 |
| 4-itemset | {D,G, H,S} | 108 |

**Table 5:  Customer Transaction table**

| Trans id | Customer_id | Customer_name | Customer_type | Transaction_Profit |
|---|---|---|---|---|
| 1 | c01 | CA | General | 39 |
| 2 | c02 | CB | General | 44 |
| 3 | c03 | CC | General | 30 |
| 4 | c04 | CD | **Premium** | **47** |
| 5 | c05 | CE | General | 21 |
| 6 | c06 | CF | General | 10 |
| 7 | c07 | CG | General | 13 |
| 8 | c08 | CH | General | 35 |
| 9 | c09 | CI | **Premium** | **49** |
| 10 | c10 | CJ | **Premium** | **49** |
| 11 | c11 | CK | General | 31 |
| 12 | c12 | CL | General | 40 |
| 13 | c13 | CM | General | 23 |
| 14 | c14 | CN | **Premium** | **71** |
| 15 | c15 | CO | General | 31 |
| 16 | c16 | CJ | General | 2 |
| 17 | c17 | CK | General | 21 |
| 18 | c18 | CL | General | 41 |
| 19 | c19 | CM | General | 25 |
| 20 | c20 | CN | General | 17 |
| 21 | c21 | CO | General | 42 |
| 22 | c22 | CP | General | 13 |
| 23 | c23 | CQ | General | 32 |
| 24 | c24 | CR | General | 17 |
| 25 | c25 | CS | **Premium** | **48** |
| 26 | c26 | CT | General | 19 |
| 27 | c27 | CU | General | 29 |
| 28 | c28 | CJ | General | 30 |
| 29 | c29 | CV | **Premium** | **50** |
| 30 | c30 | CW | General | 13 |
| 31 | c31 | CX | General | 32 |
| 32 | c32 | CY | General | 13 |
| 33 | c33 | CZ | General | 38 |
| 34 | c34 | AV | General | 9 |
| 35 | c35 | AB | General | 32 |

Given a user-specified maximum support threshold maxsup, we are interested in a rule X if sup(X) < maxsup. Rare rules are those rules appearing below the maximum support value. By setting the value of maximum support threshold to 40%, the rare itemsets generated from table 1are listed in table 3. Rare itemsets of users' interest or high utility rare itemsets fall below a maximum support value but above a user provided high utility threshold. If high utility threshold is set as 45, the high utility rare itemsets generated are listed in table 4.

HURI algorithm produces high utility rare itemsetst according to users' interest. CSHURI, the modified version of HURI, is used as a base for customer utility mining. Customer utility mining aims at finding customers who purchase high profitable rare items. This type of customers gives almost hundred percent contributions towards the overall profit of the transaction. For e.g., by setting the user provided customer utility as 45, customers can be classified as Premium (Profit transaction > 45) or General Customer (table 5). After finding this type of customer, the customers can be categorized and accordingly gold or silver cards can be issued for this type of customers or new scheme can be introduced for them.

CSHURI algorithm, **C**ustomer **S**egmentation using **HURI,** uses two-phase HURI algorithm [6] for finding customers who purchase high utility rare itemsets, after generating high utility rare itemsets. In CSHURI Algorithm (Figure 1), customer classification is done through three phases:-

(i) In first phase, rare itemsets are generated by considering those itemsets which have support value less than the maximum support threshold (e.g. table 3).

(ii) In second phase, by inputting the utility threshold value according to users' interest, rare itemsets having utility value greater than the minimum utility threshold are generated (e.g. table 4).

(iii) Finally in the last phase, customers are classified according the type of items being purchased. By setting the customer utility threshold, all transactions having profit greater than customer utility threshold are found and accordingly customers of corresponding transactions are classified (e.g. table 5).

## 4. CONCLUSIONS AND FUTURE SCOPE

In retail markets, the customized service is the most crucial phase. Customer base is the prime objective of customer relationship management. The key to attain this objective is to understand the buying behaviour of customers. Clear Customer understanding requires properly focused customer segmentation and actions to maximize customer retention, loyalty and profitability. Data mining techniques are nowadays used to predict buying behaviour of customers by analyzing large amounts of customer and transaction data. Stores can use its customer base as a bargaining power to strike discount deals. Customer Utility mining can generate insights, which can lead to effective customer segmentation in retail marketing. The proposed CSHURI algorithm, **C**ustomer **S**egmentation using **HURI,** finds customers purchasing high utility rare itemsets and accordingly customers are segmented easily. These customers provide maximum profit to the overall transaction scenario. Marketers can then make business strategies by providing special services to premium customers to delight the customers and retain the customer with same business. Premium customers can be given special privileges, credit facilities to customers, special cards, discount offer, promotion packages.

Enterprises can also use data mining to minimize purchasing costs; score suppliers by rating the quality of their goods and services; identify the most effective promotions; identify profitable itemsets. Also after identification of high utility rare itemsets, marketers can do the promotion or advertising of such itemsets to increase the overall profit of the business. The association of customers with different products can be found with the help of data mining systems. The knowledge generated from CSHURI will be useful for retail businesses in decision-making process. The outcome of CSHURI would enable the top management or business analyst in crucial decision-making such as catalog design, providing credit facility, cross marketing, finalizing discount policy, analyzing consumers' buying behaviour, organizing shelf space, loss-leader analysis and quality improvement in supermarket.

The future work includes more modified versions of HURI by incorporating temporal and fuzzy concept for finding those rare items, which provide maximum profit to a transaction. The future work also includes finding the share of profitable transactions in the whole business, which may help in effective planning for retail marketing in Supermarket and online stream mining.

## Authors

**Mrs. Jyothi Pillai** is Associate Professor in Department of Computer Applications at Bhilai Institute of Technology, Durg (C.G.), India. She is a post-graduate from Barkatullah University, India. She is a Life member of Indian Society for Technical Education. She has a total teaching experience of 16½ years. She has a total of 15 Research papers published in National / International Journals / Conferences into her credit. Presently, she is pursuing Ph.D. from Pt. Ravi Shankar Shukla University, Raipur under the guidance of Dr. O.P.Vyas, IIIT, Allahabad.

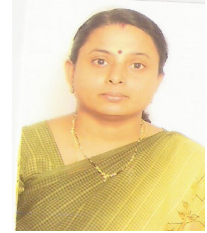

**Dr.O.P.Vyas** is currently working as Professor and Incharge Officer (Doctoral Research Section) in Indian Institute of Information Technology-Allahabad (Govt. of India's Center of Excellence in I.T.). Dr.Vyas has done M.Tech.(Computer Science) from IIT Kharagpur and has done Ph.D. work in joint collaboration with Technical University of Kaiserslautern (Germany) and I.I.T.Kharagpur. With more than 25 years of academic experience Dr.Vyas has guided Four Scholars for the successful award of Ph.D. degree and has more than 80 research publications with two books to his credit. His current research interests are Linked Data Mining and Service Oriented Architectures.

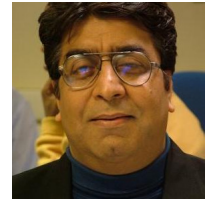